\documentclass[sn-mathphys-num]{sn-jnl}

\usepackage{graphicx}%
\usepackage{multirow}%
\usepackage{amsmath,amssymb,amsfonts}%
\usepackage{amsthm}%
\usepackage{mathrsfs}%
\usepackage[title]{appendix}%
\usepackage{xcolor}%
\usepackage{textcomp}%
\usepackage{manyfoot}%
\usepackage{booktabs}%
\usepackage{algorithm}%
\usepackage{algorithmicx}%
\usepackage{algpseudocode}%
\usepackage{listings}%
\usepackage{url}
\usepackage{amsmath, amsthm, amsfonts, enumerate}
\usepackage{comment}

\theoremstyle{thmstyleone}%
\newtheorem{theorem}{Theorem}
\newtheorem{proposition}[theorem]{Proposition}%

\theoremstyle{thmstyletwo}%
\newtheorem{example}{Example}%

\theoremstyle{thmstylethree}%
\newtheorem{definition}{Definition}%

\begin{document}

\title[The Negative Participation Paradox in Three-Candidate Instant Runoff Elections]{The Negative Participation Paradox in Three-Candidate Instant Runoff Elections}


\author[1]{\fnm{David} \sur{McCune}}\email{mccuned@william.jewell.edu}

\author[2]{\fnm{Jennifer} \sur{Wilson}}\email{WilsonJ@newschool.edu}

\affil*[1]{\orgdiv{Department of Mathematics and Data Science}, \orgname{William Jewel College}, \orgaddress{\street{500 College Hill}, \city{Liberty}, \postcode{64068}, \state{MO}, \country{United States}}}

\affil[2]{\orgdiv{Department of Natural Sciences and Mathematics}, \orgname{The New School}, \orgaddress{\street{66 West 12th St}, \city{New York}, \postcode{10011}, \state{NY}, \country{United States}}}


\abstract{We provide theoretical and empirical  estimates for the likelihood of a negative participation paradox under instant runoff voting in three-candidate elections.  We determine the probability of the paradox and  related conditional probabilities based on the impartial anonymous culture and impartial culture models for both complete and partial ballots. We compare these results to the empirical likelihood of a negative participation paradox occurring under instant runoff voting in a large database of voter profiles which have been reduced to three candidates. Lastly, we analyze the relative likelihood of this paradox  in comparison to other well-known paradoxes. }




\maketitle

\newpage
\section{Introduction}

The voting method of instant runoff voting (IRV) is famously susceptible to many classical paradoxes. For example, under IRV an election can demonstrate an upward or downward monotonicity paradox (Fishburn \& Brams, 1983;  McCune and Graham-Squire, 2023; Miller, 2017; Ornstein \& Norman, 2014), or a no-show paradox (Felsenthal \& Tideman, 2013; Fishburn \& Brams, 1983; Moulin, 1988). IRV is also susceptible to paradoxes of negative participation, a lesser-studied type of paradox. The purpose of this article is to examine the likelihood that a three-candidate IRV election demonstrates this paradox.

An election is said to demonstrate a negative participation paradox under a given voting method if the addition of voters who rank  candidate $A$ last causes $A$ to go from losing to winning. We use the term ``negative participation paradox'' following (Kamwa et al., 2023), but other authors use language such as ``violation of negative involvement'' (P{\'e}rez, 2001). This paradox is closely related to the notion of a positive participation paradox (Kamwa et al., 2023), also referred to in the language of ``positive involvement'' (Saari, 1995), where the addition of voters who rank candidate $A$ uniquely first causes $A$ to go from winning to losing. Such outcomes are sometimes referred to as strong no-show paradoxes (Duddy, 2014; P\'{e}rez,  2001).  To the best of our knowledge, positive participation paradoxes have received more attention in prior literature than negative participation paradoxes, even though IRV is not susceptible to the positive type. Prior research on positive participation paradoxes focuses on other voting methods such as Condorcet methods (Duddy, 2014; Holliday, 2024; Holliday \& Pacuit, 2021; P\'{e}rez, 2001); this literature tends to use the term ``strong no-show paradox'' instead of positive participation paradox. Much of the prior work concerning negative participation paradoxes also examines the susceptibility of Condorcet methods to this paradox (Duddy, 2014; P\'{e}rez 2001), while some focuses on scoring runoff rules like IRV (Kamwa et al., 2023, Lepelley \& Merlin, 2001). Much previous work on these kinds of voting paradoxes focuses on elections with a small number of candidates, often analyzing elections with only three candidates. Our work similarly focuses on the three-candidate case.

This paper makes one small and two substantive contributions regarding the likelihood that an election demonstrates a negative participation paradox under IRV in a three-candidate election. First, for our small contribution we provide new theoretical results under the impartial anonymous culture (IAC) model  of voter behavior. Most of the interesting work in this vein has been done in (Kamwa et al., 2023, Lepelley \& Merlin, 2001), and our results in this area fill some small gaps left from those papers. Second, we provide results under the impartial culture (IC) model of voter behavior. We are able to obtain these new results using geometric techniques developed in (Kamwa \& Merlin, 2015) and (Saari \& Tataru, 1999). Third, we provide empirical results using a large database of IRV elections from American and Scottish political elections, as well as elections from the American Psychological Association. Ours is the first study which provides such empirical results; the only other previous work which searched for negative participation paradoxes in real-world ballot data is (McCune, 2024), which focuses on multiwinner single transferable vote elections from Scotland. Our general finding is that negative participation paradoxes occur at a much higher rate than other paradoxes.

Because we are interested in providing both theoretical and empirical analyses of the negative participation paradox we consider situations when all voters provide complete rankings and when voters are allowed to provide only partial rankings. This allows for a better comparison between  theoretical and empirical results since partial ballots are extremely common in real-world election data. On the theoretical side, we provide estimates that the likelihood of a negative participation paradox occurs under IRV for three candidates under under both the impartial anonymous culture (IAC)  and impartial culture (IC) models. For the IAC model, we analyze the complete and partial ballot cases using the software package Normaliz and Monte Carlo simulations, respectively.  For the IC model we use geometric techniques for both full and partial ballots. 

The widespread use of the method of single transferable vote in Scotland and other locations, as well as the recent increase in the use of IRV in the US, has made available a large collection of preference ballot data from real-world elections. When an election contained more than three candidates,  we reduce each election to three candidates by running a number of initial rounds of IRV.  We then determine the number of elections susceptible to a negative participation paradox using the existing ballot data as well as data which we make more ``complete'' using numerical techniques which we describe in Section \ref{empirical}. 

The paper is organized as follows. Section \ref{preliminaries} provides  relevant definitions as well as a motivating example. In Section \ref{analytical} we provide theoretical results under the IAC and IC models of voter behavior. As mentioned above, our primary contribution is for the IC model, as the IAC model has been mostly analyzed previously.  
In Section \ref{empirical} we give results using a large database of real-world elections. Section \ref{comparison} provides a discussion and comparison of the likelihood of negative participation paradoxes to other well-known paradoxes. We conclude in Section \ref{conclusion}.

\section{Preliminaries}\label{preliminaries}
Instant-runoff voting (IRV), often colloquially referred to as ``ranked-choice voting'', has become increasingly popular in the United States as an alternative to the method of plurality. In an election which uses IRV voters cast preference ballots with a (possibly partial) linear ranking of the candidates. After an election, the ballots are aggregated into a preference profile, which shows the number of each type of ballot cast. For example, Table \ref{ward8} shows a preference profile involving the three candidates $J$, $S$, and $W$. The number 1430 denotes that 1430 voters ranked $J$ first, $S$ second, and $W$ third. We use $\succ$ to denote when a voter ranks one candidate over another, so that 1430 voters cast the ballot $J\succ S \succ W$. An IRV election proceeds in rounds. In each round, the number of first-place votes for each candidate is calculated. If a candidate has a majority of the (remaining) first-place votes, they are declared the winner. If no candidate receives a majority of votes, the candidate with the fewest first-place votes is eliminated and the names of the remaining candidates on the affected ballots are moved up. The process repeats until a candidate is declared the winner.\footnote{The issue of ties does not arise in our work in this paper, and thus we ignore the possibility of multiple winners.} In the case when partial ballots are allowed, any ballot in which all candidates have been eliminated is considered ``exhausted'' and plays no further role in the determination of the winner. 

When there are only three candidates, at most two rounds are required to determine a winner; in general, several rounds may be required. In our analysis of the real-world database  in Section \ref{empirical}, we run a sufficient number of rounds for each set of voter preferences until the number of candidates is reduced to three and consider the resulting three-candidate election. We demonstrate IRV in such a case in the following example.

\begin{example} The 2003 election for Minneapolis City Council in Ward 8 involved  four not write-in candidates: Andrea Jenkins, Soren Stevenson, Bob Sullentrop, and Terry White. Sullentrop earned the fewest first-place votes and is eliminated first, creating the three-candidate preference profile shown in Table \ref{ward8}. The resulting first-place vote totals for Jenkins, Stevenson, and White are 3591, 3640, and 668, respectively. White is eliminated in the next round, causing a transfer of 303 votes to Jenkins and 216 to Stevenson. As a result, Jenkins defeats Stevenson 3894 votes to 3856.

\begin{table} \label{ward8}
\begin{tabular}{l|ccccccccc}
Num. Voters & 1149 & 1430 &1012 & 830& 1989 &821&149&303&216\\
\hline
1st choice & $J$ & $J$ & $J$ & $S$ & $S$ & $S$ & $W$ & $W$ & $W$\\
2nd choice & & $S$ & $W$ & & $J$ & $W$ & & $J$ & $S$ \\
3rd choice &  & $W$ & $S$ & & $W$ & $J$ & & $S$ & $J$\\
\end{tabular}
\caption{Vote totals for the top 3 candidates in the 2023 election for Minneapolis City Council Ward 8}
\end{table}

\end{example}

It is well-known that IRV can produce strange outcomes when ballots are added or removed from an election (Felsenthal \& Tideman, 2013; Fishburn \& Brams, 1983; Graham-Squire, 2024; Graham-Squire and McCune, 2023; Kamwa et al., 2023; Lepelley \& Merlin, 2001; McCune \& Graham-Squire, 2023), or when candidates are added or removed from an election (Green-Armytage et al., 2016; McCune \& Wilson, 2023; Miller, 2019; Tideman, 1987). The example below illustrates a  type of paradoxical outcome which can occur when ballots are added to an election.

\begin{example}\label{ward8b}  Suppose we add 3000 ballots of the form $W\succ J \succ S$ to the ballots from  in Table \ref{ward8}. What effect should this have on the electoral outcome? Since these voters all prefer $W$, it would make sense for White to win after receiving this large boost of support. Since these voters prefer Jenkins to Stevenson, it would also make sense for these ballots to bolster Jenkins' victory. However, the addition of these ballots turns Stevenson into the winner.  To see this, note that with these ballots added to the electorate the initial first-place vote totals are 3591, 3640, and 3668 for Jenkins, Stevenson, and White, respectively. In this modified election Jenkins is eliminated first and after the vote transfers Stevenson defeats White 5070 votes to 4680.
\end{example}

Thus, adding thousands of ballots in which Stevenson is ranked last can result in a good  outcome for him. Note that this outcome is also true in the original four-candidate election including Sullentrop ($Su)$:  the same result is obtained if we add 3000  ballots of the form $W \succ J \succ Su \succ S$ or 3000 ballots of the form $W\succ Su \succ J \succ S$. That is, when we add ballots to create a negative participation paradox in an IRV election which has been reduced to three candidates, only the first and last rankings matter. We can fill in the intermediate rankings in an arbitrary fashion; the original election with all candidates also demonstrates a negative participation paradox as long as sufficient votes are added so that the original three-candidate plurality loser remains until eliminated in the penultimate round.

This example motivates the primary definition of the paper.

\begin{definition}
An election with preference profile $P$ is said to demonstrate a \textbf{negative participation paradox} if there exists a losing candidate $A$ and a set of identical ballots $\mathcal{B}$ with $A$ ranked last such that if we add the ballots from $\mathcal{B}$ to the ballots from $P$, candidate $A$ wins the resulting election.
\end{definition}

That is, an election demonstrates this paradox when a losing candidate can be made into a winner by adding ballots on which the loser is ranked last. We use the language of ``negative participation paradox'' following (Kamwa et al., 2023), but, as noted in the Introduction, there is no standard vocabulary around the topic. For example, much of the literature concerning these paradoxes and Condorcet methods (Holliday, 2024; Holliday \& Pacuit, 2021; P\'{e}rez, 2001) uses the the language of strong no-show paradoxes and negative and positive involvement. We note that for some researchers, a negative participation paradox can be considered a special case of a no-show paradox, a paradox briefly discussed in Section \ref{comparison}. This is not the case for us, as we use the definition of no-show paradox (sometimes also referred to as an abstention paradox) found in (Graham-Squire, 2024) and (Kamwa et al., 2023). In these articles, a three-candidate election demonstrates a no-show paradox if there exists a set of ballots $\mathcal{B}$ with the election winner $A$ ranked last such that if we remove these ballots then $A$ loses in the resulting election.

Example \ref{ward8b} shows that IRV is susceptible to negative participation paradoxes; many other voting methods such as positional scoring rules cannot exhibit this type of paradox. We also note that other researchers have studied the notion of a \emph{positive participation paradox}, which occurs if there exists a winning candidate $A$ and a set of ballots $\mathcal{B}$ ranked first such that the addition of these ballots to the electorate turns $A$ into a losing candidate. We do not analyze positive participation paradoxes in this article because IRV is not susceptible to it; however, the voting method of single transferable vote, which is the multiwinner version of IRV, is susceptible to positive participation paradoxes (see (McCune, 2024) for two real-world examples).

In the US, the most common voting method for political elections is the method of plurality, in which the candidate with the most first-place votes wins the election. Throughout the article, we refer to the candidate who receives the most first-place votes in a three-candidate election as the \emph{plurality winner}. For example, Stevenson is the plurality winner of the preference profile in Table \ref{ward8}. The \emph{plurality loser} is the candidate with the fewest first-place votes. In the Minneapolis Ward 8 election Sullentrop was the actual plurality loser; because we care only about the three-candidate case, we ignore Sullentrop and say that White is the plurality loser of the election in Table \ref{ward8}.

In a three-candidate election, a negative participation paradox can occur only when several conditions are satisfied.  First, the added votes cannot result in the plurality loser being made into the IRV winner: if the added votes rank the plurality loser last, that candidate will remain the plurality loser and be eliminated in the first round. Second, the IRV winner must be different from the plurality winner. To see this, suppose that the IRV winner is $A$ and the candidates are ranked by first-place votes $A > B >C.$ Since $A$ is the IRV winner,  $B$ must be the losing candidate who can be made into a winner by the addition of votes ranking $B$ last. But $B$ cannot be the winner in such a fashion: if $B$ becomes the new plurality loser then they are eliminated in the first round, and if $C$ remains the plurality loser than $C$ is eliminated in the first round, leaving $B$ to be eliminated in the second round as they were in the original election (the added votes increase the margin of $A$ over $B$). Hence the IRV and plurality winners must be distinct, implying it is the plurality winner who can be made to win under IRV by the addition of votes ranking them last.  Similar reasoning implies that the new votes must rank the plurality loser first so that  the original IRV winner is eliminated in the first round of the new election.

We use this logic to determine the probability of a negative participation paradox in three-candidate elections under two models of voter behavior in the next section.

\section{The Impartial Anonymous and Impartial  Culture Models}\label{analytical}

In this section we provide probabilities that an election demonstrates a negative participation paradox under the impartial anonymous culture (IAC) and impartial culture (IC) models of voter behavior.  Suppose the three candidates are labeled $A$, $B$ and $C$, and there are $V$ voters. If voters must provide complete rankings, there are six possible candidate rankings; if voters can provide partial rankings, this number increases to 9.  Assume  the preference profiles for complete and partial rankings are  as shown in Tables \ref{profile_complete_ballots} and \ref{profile_partial_ballots} where in each case the numbers $a_i, b_i$, and $c_i$ indicate the number of voters with the corresponding ranking and that these numbers sum to  $V.$ Note in Table  \ref{profile_partial_ballots} we equate the rankings $A \succ B$ and $A \succ B \succ C$ and hence do not list them separately. For either model of voter behavior, we are concerned with the \emph{limiting probability} that an election demonstrates a negative participation paradox, where $V \rightarrow \infty$.

Before providing the limiting probabilities, we examine the conditions (expressed as inequalities) under which an election demonstrates a negative participation paradox.

\begin{table}[tbh]
\centering
\begin{tabular}{cccccc}
$a_1$ & $a_2$ & $b_1$ & $b_2$ & $c_1$ & $c_2$ \\
\hline
A  &A & B & B & C & C \\
B  &C & A & C & A& B \\
C  &B & C & A & B & A \\
 
\end{tabular}
\caption{Preference profile with complete preferences.}
\label{profile_complete_ballots}

\end{table}

\begin{table}[tbh]

\centering
\begin{tabular}{ccccccccc}
$a_0$ & $a_1$ & $a_2$ & $b_0$ & $b_1$ & $b_2$ & $c_0$& $c_1$ & $c_2$ \\
\hline
A& A &A &B & B & B &C&C&  C \\
& B&C   && A & C &&A  &B \\
& C &B  && C & A &&B&  A \\
\end{tabular}
\caption{Preference profile with some partial preferences.}
\label{profile_partial_ballots}
\end{table}

In the complete ballot case, we can assume without loss of generality that
\begin{equation}\label{oldorder}
 a_1+a_2>b_1+b_2>c_1+c_2;
 \end{equation} 
 similarly, in the partial ballot case we assume without loss of generality that 
 \begin{equation}\label{partial_oldorder}
 a_0+a_1+a_2>b_0+b_1+b_2>c_0+c_1+c_2.
 \end{equation}
 
That is, we assume $A$ is the unique plurality winner and $C$ is the unique plurality loser. As described in Section \ref{preliminaries},  a negative participation paradox can only occur when the plurality ($A$) and IRV  ($B$) winners are distinct and when additional votes are added with the ranking $C \succ B \succ A.$  Hence, initially
\begin{equation}\label{oldwinner}
a_1+a_2+c_1 < b_1+b_2+c_2.
\end{equation}
 
If $z$ votes are added, then we require  
$B$ to be eliminated in the first round followed by $C$ in the second round. Hence,
\begin{equation*}\label{newwinner}
b_1+b_2< c_1+c_2+ z,  \quad a_1+a_2+b_1 > b_2+c_1+c_2+z. 
\end{equation*}
Solving for $z$ in each of these yields  for $z$ yields  
\begin{equation}\label{zinterval}b_1+b_2-(c_1+c_2)< z< a_1+a_2+b_1 -(b_2+c_1+c_2)\end{equation}
which implies
\begin{equation}\label{zcond}
b_1+b_2-(c_1+c_2) < a_1+a_2+b_1 -(b_2+c_1+c_2) \quad \mbox{ or } \quad a_1+a_2>2b_2.
\end{equation} 
It is easily seen that (\ref{oldwinner}) and (\ref{zcond}) are also sufficient to ensure a negative participation paradox can occur given that $A$ is the plurality winner and $C$ is the plurality loser\footnote{In the non-limiting case where the number of voters is finite, (\ref{zinterval}) must have room for at least one integer value of $z$ to fit in the interval, and thus in this case  (\ref{zcond}) becomes $a_1+a_2>2b_2+1$.}. Moreover, (\ref{zinterval}) provides an explicit formula for the range of added votes that result in the paradox.

In the partial rankings case, (\ref{oldwinner}) and (\ref{zcond}) are replaced by
\begin{equation}\label{partial_oldwinner}
b_0+b_1+b_2+c_2 > a_0+a_1+a_2+c_1
\end{equation}
and we obtain
\begin{equation}\label{partial_zcond}
a_0+a_1+a_2>b_0+2b_2.
\end{equation} Inequality (\ref{zinterval}) is replaced by
\begin{equation}\label{partial_zinterval} b_0+b_1+b_2-(c_0+c_1+c_2)< z< a_0+ a_1+a_2+b_1 -(b_2+c_0+c_1+c_2).
\end{equation}

Before defining and analyzing the two models, we note that they are widely used in the social choice literature to provide \emph{a priori} estimates for the probabilities of many phenomena in social choice, and they tend to provide theoretical upper bounds for such probabilities. For example, in three-candidate elections the probability that an election fails to contain Condorcet winner is 6.25\% under the IAC model (this is easily calculated using Normaliz); the probability falls essentially to 0 in real-world elections (Graham-Squire \& McCune, 2023; Myers, 2024; Song, 2023). Thus, the models are useful for providing theoretical ``worst-case'' estimates for various election probabilities.

\subsection{Impartial Anonymous Culture}

Under the impartial anonymous culture (IAC) model, the vector of vote totals at the top of the preference profile is assumed to be chosen at random. For example, in the complete ballot case we choose at random a vector of the form $(a_1,a_2,b_1,b_2,c_1,c_2)$ subject to the constraints that each entry is non-negative and the six vote totals sum to the number of voters. Since we examine only limiting probabilities, we choose a vector of proportions of length six in the complete ballot case or a vector of proportions of length nine in the partial ballot case. Limiting probabilities involving  IAC can be determined using  the theory of Ehrhart polynomials, which has been implemented in the software package Normaliz (Bruns et al., 2022). The probabilities presented in Proposition \ref{IAC_prop} were calculated using Normaliz. The  probability of a negative participation paradox using IRV assuming complete rankings appears in (Kamwa et al., 2023). In the following proposition, we include this result along with related conditional probabilities for reference. 

Assuming, as in the previous section, that the first-place vote ranking is $A >B>C,$ the probability that  an election demonstrates a negative participation paradox is equal to the probability that (\ref{oldwinner}) and (\ref{zcond}) are satisfied. This probability, shown in (i), is obtained directly from (Kamwa et al., 2023). The results in (ii) and (iii) below are conditional probabilities, obtained directly from Normaliz. If an election demonstrates a negative participation paradox then there cannot be a majority candidate (a candidate who receives a majority of first-place votes) because the plurality and IRV winners would coincide. Thus, in (ii),  the probability that  an election demonstrates a negative participation paradox winner assuming the election does not contain a majority candidate is equal to the the probability of a negative participation paradox divided by the probability that an election does not have a majority candidate. Likewise, in (iii), the  probability that an election demonstrates a negative participation paradox assuming the IRV winner is not the plurality winner is equal to the  probability that an election demonstrates a negative participation paradox divided by  the probability the IRV winner is not the plurality winner. Previous studies such as (Miller, 2017) have shown that as an election gets ``closer'' the probability of observing various voting paradoxes increases; items (ii) and (iii) of the proposition also demonstrate this for negative participation paradoxes.

\begin{proposition}\label{IAC_prop}
In a 3-candidate election in which all voters provide complete preferences, under IAC

\begin{enumerate}[i]
\item the limiting probability that an election demonstrates a negative participation paradox  is $7/96 = 7.29\%$ (Kamwa et al., 2023).
\item the limiting probability that an election demonstrates a negative participation paradox  is $1/6 = 16.67\%$, assuming the election does not contain a majority candidate.
\item the limiting probability that an election demonstrates a negative participation paradox is $42/71 = 59.15\%$, assuming the IRV winner is not the plurality winner.

\end{enumerate}

\end{proposition}

\vspace{.05 in}
Normaliz also allows us to obtain the liminting probabilities under IAC for the partial ballot case.

\vspace{.05 in}

\begin{proposition}\label{IAC_prop_partial}
In a 3-candidate election in which voters can provide partial preferences, under IAC

\begin{enumerate}[i]
\item the limiting probability that an election demonstrates a negative participation paradox  is $169/2304 = 7.34\%$.
\item the limiting probability that an election demonstrates a negative participation paradox  is $169/1305 = 12.95\%$, assuming the election does not contain a majority candidate.
\item the limiting probability that an election demonstrates a negative participation paradox is $117/187 = 62.57\%$, assuming the IRV winner is not the plurality winner.

\end{enumerate}

\end{proposition}


\subsection{Impartial  Culture}

Under the impartial  Culture (IC)  model, each voter selects uniformly and  independently among all possible rankings of the candidates. We can approximate the likelihood of a negative participation paradox for a large number of voters by calculating the limiting probability as $V\rightarrow \infty$ using the geometric methods introduced in (Saari \& Tataru, 1999) and subsequently elaborated on in (Kamwa \& Merlin, 2015). The probability that there is a majority candidate is zero in the limiting case under the IC model, and thus we provide counterpoints to only two of the probabilities identified in Proposition \ref{IAC_prop}.

\begin{proposition}\label{IC_prop}
In a 3-candidate election in which all voters provide complete preferences, under IC

\begin{enumerate}[i]
\item the limiting probability that an election demonstrates a negative participation paradox  is approximately $0.157737 \approx 15.77\%$.
\item the limiting probability that an election demonstrates a negative participation paradox  is   $0.644717 \approx 64.47\%$,  assuming the IRV winner is not the plurality winner.

\end{enumerate}

\end{proposition}

\begin{proposition}\label{IC_proppartial}
In a 3-candidate election in which voters provide partial preferences, under IC

\begin{enumerate}[i]
\item the limiting probability that an election demonstrates a negative participation paradox  is  approximately $0.143861 \approx 14.39\%$. 
\item the limiting probability that an election demonstrates a negative participation paradox  is approximately $0.672803 \approx 67.28\%$, assuming the IRV winner is not the plurality winner.

\end{enumerate}

\end{proposition}

\vskip 2mm

The proofs of both propositions can be found in the Appendix.


 \subsection{Analysis of IAC and IC Results}
 
 Overall,  a negative participation paradox is twice as likely  under the IC than  the IAC model, and much more likely assuming the IRV and plurality winners are distinct.  This is to be expected, as the IAC model frequently produces preference profiles containing a majority candidate while this is impossible under IC in the limiting case. Under the IC model, the probability when using complete preferences is slightly higher than when using partial preferences; we investigate why this should also be the case in real-world elections in Section \ref{empirical}. Under IAC, the difference between partial and complete preferences for the unconditioned probability is negligible. 

 Under the IAC model, the probability that an election does not contain a majority candidate is $7/16=43.75\%$ in the complete preference case and $145/256=56.64\%$ in the partial ballot case. Thus, even though the unconditioned limiting probabilities are approximately equal, the probability conditioned on the absence of a majority candidate is lower in the partial ballot case.
 
 Under both models, the probability for the case of partial preferences is larger than for complete preferences when we assume the IRV and plurality winners are distinct. This is possibly because the probability that these winners are different is somewhat lower for partial preferences than for complete preferences (in the IC case, 21\% versus 24\%). With partial ballots it is less likely that a candidate ranking second in first-place votes will gain sufficient additional votes, after the plurality loser is eliminated, to overcome the vote total of the plurality winner.   

\begin{figure}[tbh]
    \centering
    \begin{tabular}{ccc}
    \includegraphics[width=55mm,height=60.5mm]{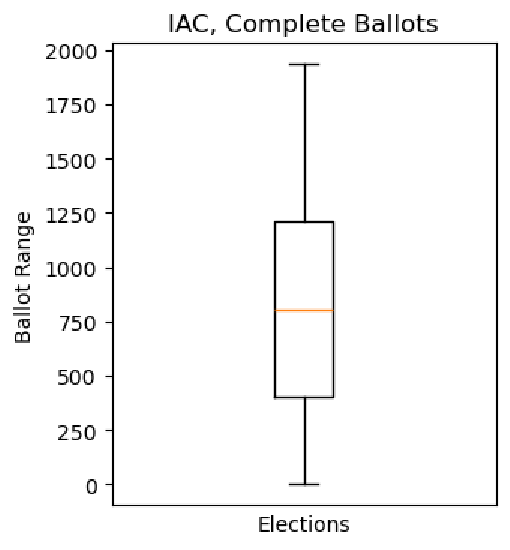} & \includegraphics[width=55mm]{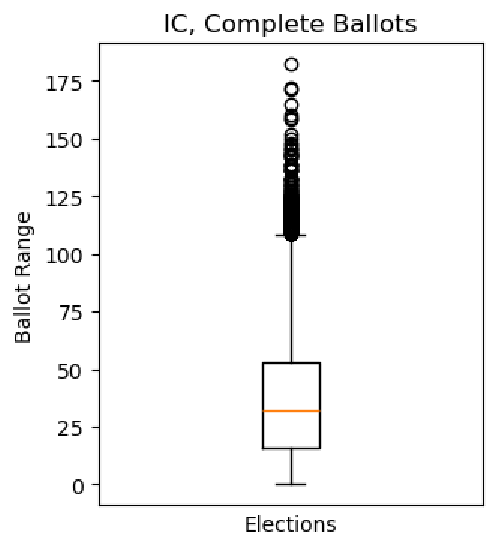}
    
    \end{tabular}
    \caption{The ballot ranges of elections which demonstrates a paradox under the IAC and IC models with 3901 voters.}
    \label{boxplots}
\end{figure}

In addition to calculating probabilities under each model, we also examine the range of the possible number of ballots (given by Inequalities \ref{zinterval} and \ref{partial_zinterval}) which can be added to demonstrate a paradox. If an election exhibits a paradox, we say the \emph{ballot range} of the election is the maximum number of ballots which can be added to demonstrate the paradox minus the minimum number of ballots. In Example \ref{ward8b}, we can add as many as 3389 ballots to demonstrate the paradox and as few as 2924, and thus the ballot range for that election is $3389-2924=465$. The range of possible number of ballots gives a measure of how ``close'' the election is to one that does not demonstrate the paradox---the larger the range, the more susceptible the election.

 To investigate such ballot ranges, for each model we generated 100,000 elections with complete ballots using 3901 voters. For each election which demonstrated a negative participation paradox we calculated the ballot range; the results are displayed in Figure \ref{boxplots} for elections with complete ballots. We also generated box plots for elections with partial ballots but the corresponding plots are not markedly different. The left box plot for IAC uses the 7268 generated elections which demonstrated a paradox under this model; the box plot for IC uses 15128 elections.\footnote{Note that 15128 out of 100,000 runs gives an estimated probability a bit far from the probability of 15.77\% given in Proposition \ref{IC_prop}. The reason is that even with a large electorate of 3901 voters, approximately 2400 simulated elections contain at least two candidates with the same number of first-place votes. If we ignore such elections with ties (which is the case for a limiting probability) then our estimated probability becomes much closer to 15.77\%.} The median ballot range under the IC model is 32 and the maximum is 182. Thus, relative to an electorate size of 3901, under IC elections which generate a paradox tend to have small ballot ranges. In particular, 227 of the generated elections which demonstrate a paradox have a ballot range of zero, and the paradoxical outcome is as ``close'' as possible to an election which does not exhibit a paradox. Under the IAC model the median ballot range is 806 and the maximum is 1934. Thus, while IAC produces far fewer paradoxes than IC, this model tends to produce much larger ballot ranges. In this case, only two of the generated elections produce a ballot range of zero.

  We compare our IC and IAC results with an empirical analysis in the next section.

  \section{Empirical Results}\label{empirical}

In this section we present empirical results using a large database of real-world IRV elections. Our data comes from three sources. First, we use single-winner political elections from the United States with at least three (not write-in) candidates. This data is available at the FairVote data repository (Otis, 2022), and contains ballot data from municipal elections in cities such as San Francisco, CA and Minneapolis, MN. The repository also contains ballot data from federal elections in Alaska and Maine. Second, we use single-winner elections from the American Psychological Association, which generally is willing to share ballot data for ranked-choice presidential elections and elections for the Board of Directors. Some of this data is available on preflib.org (Mattei \& Walsh, 2013), and some was shared directly with the first author. Third, we use single-winner political elections from Scotland from a database of Scottish local government elections. The ballot data for these elections is available at \url{https://github.com/mggg/scot-elex}.
 
 In total, we have access to the preference profiles for 361 single-winner IRV elections with at least three candidates. Many of these elections contain more than three candidates; because we focus on the three-candidate case, for each election we run the IRV algorithm until three candidates remain and use the resulting profile. As with Example \ref{ward8}, any paradox we find in the resulting three-candidate profile will also be demonstrated in the original election. The reason is that when we create the ballots which produce the paradox only the first and last rankings matter; the intermediate rankings can be filled in arbitrarily using the remaining candidates from the original candidate set. Of course, it is possible that an election with four or more candidates demonstrates a paradox which cannot be found when we reduce down to three candidates; our methodology would not find such elections. We note also that in trying to characterize the likelihood of such a paradox in real settings, we are assuming that voters' relative preferences over the remaining three candidates in each election would not change had those candidates been the only ones running. Such an assumption seems reasonable given the rationale of the IRV algorithm itself, which sequentially eliminates candidates and bumps up remaining candidates in each voter's preference order.

In Section \ref{analytical} we analyzed the likelihood of negative participation paradoxes in two cases: when all voters provide complete preferences and when some voters cast partial ballots. To provide a similar empirical analysis, with our real data we also give results using fully complete preferences and using partial ballots. In real elections voters often provide partial rankings of the candidates, and thus our results from partial ballots represent the ``actual'' results, the results obtained from using the actual ballot data. To create complete preferences, we proportionally fill in partial ballots using the available complete ballots to create hypothetical results for the case that all voters submit complete preferences. 

To see how we fill in the ballots, consider the preferences profile from Table \ref{ward8}. We need to take the bullet votes for $J$ and extend them to complete ballots over the three candidates. Since $1430+1012=2442$ voters rank $J$ first and provide complete rankings, we extend $1430/2442*100\%$ of the 1149 bullet votes for $J$ to ballots of the form $J\succ S \succ W$ and $1012/2442*100\%$ of the bullet votes for $J$ to ballots of the form $J\succ W\succ S$, rounding to the nearest integer. As a result, 673 of the 1149 bullet votes are extended to  $J\succ S \succ W$ and 476 are extended to  $J\succ W \succ S$. This methodology is used on every three-candidate profile to create complete preferences.

In Section \ref{analytical} we do not create complete preference data proportionally in this manner; instead, we generate complete ballots separately from partial ballots. The reason for this difference in methodology is that when using theoretical models like IAC or IC, we have no pre-existing data from which to draw, and if we treat the cases separately then we can use the geometric techniques outlined in Section \ref{analytical} to calculate probabilities for both cases. If we were to create complete ballots proportionally after first using IAC or IC to generate partial ballots, the complete ballot probabilities are similar but not equal to the probabilities in Propositions \ref{IAC_prop} and \ref{IC_prop}. (We confirmed this using simulations.) Thus, a comparison between our theoretical and empirical results for complete ballots is less useful than a comparison between our results for partial ballots, although both comparisons are still valuable.

 \begin{table}[tbh]
 \begin{tabular}{c|c|c|c|c}
 Jurisdiction& USA & APA & Scotland& Total\\
 \hline
Elections & 304 & 27 & 30& 361\\
 \hline
 No Majority Cand & 191 & 23 & 30& 244\\
 \hline
 IRV Winner $\neq$ Plurality Winner & 18, 30& 3, 3 & 5, 8& 26, 41\\
 \hline
 Paradoxes & 7, 17 & 3, 3 & 2, 3& 12, 23\\
 \end{tabular}
 \caption{Results from real-world elections. In the final two rows, the first number is obtained using the actual ballots and the second is obtained from ballots which have been completed.}
 \label{empirical_results1}
 \end{table}
 

   \begin{table}[tbh]
 \begin{tabular}{c|c|c}
 & Partial (actual) & Completed\\
 \hline
 Probability of a paradox & 3.3\%  & 6.4\%   \\
 \hline
Given No Majority Cand & 4.9\% & 9.4\%\\
 \hline
 Given  IRV Winner $\neq$ Plurality Winner & 46.2\% & 56.1\%\\
 \end{tabular}
 \caption{Probability of a negative participation paradox in real-world elections.}
 \label{empirical_prob}
 \end{table}


 Our results are summarized in Table \ref{empirical_results1}. The top row shows the number of elections from each jurisdiction and the second row shows the number of elections without a majority candidate. The third row shows the number of elections in which the IRV winner is different from the plurality winner. The first number is the number of such elections when using the actual data with partial ballots; the second number is the number of such elections when using complete preferences, created as described above. The fourth row shows the number of negative participation paradoxes, using the same format. For example, there are 191 American political elections without a majority candidate. Of these, the IRV winner is not the plurality winner in 18 (respectively 30) elections when using the actual ballot data (respectively, complete preference data). Seven American elections demonstrate a paradox when using the actual data, which rises to 17 when we complete the ballots. By summing all the instances of the paradox across all the election categories and dividing by the appropriate denominator we obtain the probabilities obtained in Table \ref{empirical_prob}. 
 
 As with the IAC and IC models, we obtain higher probabilities for complete preferences for the two conditional probabilities. Unlike those models, we obtain higher probabilities for complete preferences for the unconditioned probabilities as well.
 This is partly due to the fact that completing ballots can ``introduce'' paradoxes to elections where no potential existed in the actual voter profile.    

We illustrate this phenomenon using  the 2022 IRV mayoral election from Corvallis, Oregon.
\begin{example}\label{corvallis}
The 2022 election for mayor of Corvallis, Oregon, marked the first use of IRV for a mayoral election in the city. Ignoring write-in candidates, the election contained the three candidates Hogg ($H$), Maughan ($M$), and Struthers ($S$). The preference profile for the actual election is displayed in the top of Table \ref{corvallis_profiles}. 

In the actual election, the initial vote totals are 5535, 7917, 7147 for $H$, $M$, and $S$, respectively. After the elimination of Hogg, Maughan defeats Struthers 9893 to 9859. Note that the plurality and IRV winners agree, and thus the election cannot demonstrate a negative participation paradox.

If we proportionally complete the partial ballots then we obtain the profile in the bottom of Table \ref{corvallis_profiles}. In this modified election Hogg is still eliminated first, but Struthers defeats Maughan 10349 votes to 10250 in the last round. Note that Inequalities (\ref{oldwinner}) and (\ref{zcond}) are satisfied, and if we add $z$ ballots of the form $H\succ S \succ M$ where $z \in [2383, 4534]$ then Maughan becomes the winner.

\begin{table}[tbh]
\centering
\begin{tabular}{l|ccccccccc}
Num. Voters & 847 & 1976 & 2712 & 372 & 2380 & 5165 & 517 & 2398 & 4232\\
\hline
1st choice & $H$ & $H$ & $H$ & $M$ & $M$ & $M$ & $S$ & $S$ & $S$\\
2nd choice & & $M$ & $S$ & & $H$ & $S$ & & $H$ & $M$\\
3rd choice & & $S$ & $M$ & & $S$ & $H$ & & $M$ & $H$\\
 
\end{tabular} 

\vskip 2mm

\begin{tabular}{l|cccccc}
Num. Voters & 2333 & 3202 & 2497 & 5420 & 2585 & 4562\\
 \hline
1st choice &  $H$ & $H$ & $M$ & $M$ & $S$ & $S$\\
2nd choice &  $M$ & $S$ &  $H$ & $S$ &  $H$ & $M$\\
3rd choice &  $S$ & $M$  & $S$ & $H$ &  $M$ & $H$\\
\end{tabular}
\caption{The actual preference profile for the 2022 Corvallis mayoral election (Top) and the profile with preferences completed proportionally (Bottom).}
\label{corvallis_profiles}

\end{table}

\end{example}

Overall, the frequency with which the paradox of negative participation occurs in the real-world database is lower than that which is predicted by the IAC and IC models, which is to be expected since these models tend to provide upper bounds for the probabilities of paradoxical behavior. It is interesting, though, that some of the empirical probabilities are not significantly lower than the theoretical probabilities. Often, an empirical estimate of the probability that an election demonstrates a particular voting paradox is much lower than what is predicted by a model like IAC or IC. When we assume the IRV and plurality winners are different, the empirical probability for negative participation paradoxes is surprisingly close to the probabilities reported in Section \ref{analytical}.

 We complete our analysis by examining the numbers of ballots that can be added to create a negative participation paradox in the  twelve  elections that were  susceptible to the paradox using the partial (actual) ballots.  Table \ref{empirical_results2} provides a complete list of these elections along with the total number of voters, the number of ballots that can be added to produce the paradox (expressed as a ballot interval), and the ballot range divided by the original number of voters. For example, the third row shows that we can add anywhere from 2924 to 3389  ballots of the form $W\succ J \succ S$ to create the paradox in Example \ref{ward8}, corresponding to a range of $3389-2924 = 465$. Dividing 465 by 7899 yields 5.9\%. Note that the last election in the table just barely produces a paradox, as the paradox can be demonstrated using only 266 ballots exactly.  Recall that under the IC model, the median ballot range for generated elections demonstrating the paradox was 32 with an electorate size of 3901, which corresponds to a percentage of 0.08\%. Under the IAC model, the median range was 806, yielding 20.7\%. The ranges found in the actual elections, therefore, lie somewhere between these two extremes.  
 
   \begin{table}[tbh]
\begin{tabular}{ p{2.8in}|c|c|c}
Election & Num. Voters & Ballot Interval & Range Perc\\

\hline
2020 San Francisco Board of Supervisors District 7 & 38321 & 798-1606& 2.1\%\\
2021 Minneapolis City Council Ward 2 & 8907 & 80-224& 1.6\%\\
2023 Minneapolis City Council Ward 8 & 7899 & 2924-3389& 5.9\%\\
2010 Oakland Mayor & 113217 & 2315-3951& 1.4\%\\
2016 Oakland School Director District 5& 12950 & 2876-4325&11.2\% \\
2022 Oakland School Director District 4 & 26432 & 38-598& 2.1\%\\
2021 New York City Rep Primary District 50& 8182 & 773-830&0.7\%\\
2005 APA President & 14079 & 2275-2780& 3.6\%\\
2007 APA President & 12925&2397-2718& 2.5\%\\
2020 APA Board of Directors Race 2 & 6227 & 1093-1883&12.7\% \\
2021 Argyll Bute By-Election Isle of Bute Ward$^\dagger$ & 1804 & 26-38& 0.7\%\\
2021 Highland By-Election Aird Ward$^\dagger$ & 3321 & 266-266& 0.0\%\\

\end{tabular}
 \caption{Each election which demonstrates a negative participation paradox when using the actual ballot data. The $\dagger$ denotes that these two elections were previously found (using different methodology) in (McCune, 2024).}
 \label{empirical_results2}
 \end{table}

\section{Comparison to other paradoxes}\label{comparison}

In this section we compare the conditions and likelihood that an IRV election demonstrates a negative participation paradox to other more well-known (and well-studied) paradoxes: upward and downward monotonicity paradoxes, and no-show paradoxes (see e.g. (Fishburn \& Brams, 1983), (Miller, 2017), and (Kamwa et al., 2023) for definitions and discussions of these paradoxes). In the three-candidate complete ballot setting, for an upward (resp. downward) monotonicity paradox to occur, the plurality loser must earn at least 25\% (respectively $16\frac{2}{3}$\%) of the first-place votes (Miller, 2017). In either the complete or partial ballot setting, the plurality loser must earn at least $16\frac{2}{3}$\% of the first-place votes for a no-show paradox to occur (Graham-Squire, 2024).\footnote{We remind the reader that by ``no-show paradox'' we mean that there exists a set of voters who could achieve a more desirable electoral outcome by not casting their ballots.} By contrast, for a large enough electorate there is no such positive lower bound for the plurality loser regarding negative participation paradoxes. That is, it is possible for the plurality lower to receive an arbitrarily small amount of first-place votes and yet the election demonstrates a negative participation paradox.

To see this, consider Table \ref{sect5table}. The top row shows the percentage of the first-place vote controlled by each candidate. Suppose these percentages satisfy $p_C<p_B<p_A<0.5$. Note that $B$ is the IRV winner, but if we add enough $C\succ B \succ A$ ballots so that the adjusted percentages satisfy $p^\prime_B <p^\prime_C<p^\prime_A$ (which is possible for a large enough electorate), then $A$ wins. Thus, it is possible for the plurality loser to be arbitrarily weak and yet observe this paradox. In particular, as long as no candidate earns 50\% or more of the first-place votes, it is possible to find  a preference profile which demonstrates a negative participation paradox for any distribution of first-place votes.

 \begin{table}[tbh]

\begin{tabular}{l|c|c|c}
\% Voters & $p_A$ & $p_B$ & $p_C$\\
\hline
1st choice & $A$ & $B$ & $C$\\
2nd choice & $B$ & $A$ & $B$\\
3rd choice & $C$ & $C$ & $A$\\

\end{tabular}

\caption{A table illustrating that an election can demonstrate a negative participation paradox even if the plurality loser controls a small percentage of the first-place votes.}
 \label{sect5table}
 \end{table}


 Furthermore, to create such a paradox we can control the type and number of ballots added to the original ballot data. Thus, we expect that such paradoxes are exhibited more frequently in real-world data than other paradoxes. Prior empirical research bears this out. Using essentially the same database of American elections that we use, (Graham-Squire \& McCune, 2023) found four elections which demonstrate an upward monotonicity paradox, three which demonstrate a downward monotonicity paradox, and one which demonstrates a no-show paradox. The authors did not include the APA elections in that study, but they evaluated the APA elections using the same methodology as in (Graham-Squire \& McCune, 2023) and found no paradoxes of any kind. The study in (McCune \& Graham-Squire, 2023) included the 30 Scottish elections we evaluate as part of a much larger study of single transferable vote elections, and of these 30 elections found only one which demonstrated a paradox (this election demonstrated both an upward monotonicity paradox and a no-show paradox). In total, of the 361 elections in our database there are five documented elections which demonstrate an upward monotonicity paradox, three which demonstrate a downward monotonicity paradox, and two which demonstrate a no-show paradox. As in this paper, all of the previously documented paradoxes occurred at the ``three candidate level'' in the sense that the paradox can be demonstrated after eliminating all but the final three candidates.

Based on prior empirical work, IRV three-candidate elections demonstrate a negative participation paradox 2-6 times more frequently than other classical paradoxes, an unsurprising result given the above discussion.

\section{Conclusion}\label{conclusion}

We conclude with two final comments. First, IRV proponents sometimes claim that a strength of IRV is that the method can  choose a different (and presumably more deserving) winner than the plurality winner. This is a natural point to make, since if the two winners always coincide then there is no reason to use IRV, and arguably the plurality winner sometimes should not win an election (e.g. if the plurality winner is a Condorcet loser).  However, in some sense our work shows that this feature of IRV comes with a price: when the IRV and plurality winners differ, it is often the case that the election demonstrates a negative participation paradox. That is, when IRV ``works properly'' by choosing a more deserving candidate than the plurality winner, often the plurality winner could have been the IRV winner if only the turnout of some voters who don't support the plurality winner were increased.

Second, we make no claim regarding the normative status of negative participation paradoxes. Social choice theorists care about  paradoxes such as monotonicity and no-show paradoxes in part because such paradoxes show there is some mathematical irrationality built into the mechanics of IRV. Put simply, sometimes IRV does not behave rationally in response to changes in the ballot data, and this seems normatively undesirable. On the other hand, paradoxes in this vein are hypothetical--negative participation paradoxes especially so. In Example \ref{ward8b}, it is extremely unlikely that there were thousands of potential voters in Minneapolis' eighth ward whose favorite candidate was White and whose least favorite was Stevenson, and these voters abstained from the actual election.  Example \ref{ward8b} is built using thousands of people who probably do not exist, which perhaps ultimately says nothing about whether IRV is a ``good'' or ``appropriate'' voting method. Furthermore, while negative participation paradoxes may be normatively undesirable, IRV may have other positive features which justify its use, despite the susceptibility of IRV to various voting paradoxes.

Regardless of one's normative stance on these paradoxes, the results in this paper  contribute to the literature on voting paradoxes by providing limiting probabilities regarding negative participation paradoxes in the three-candidate case under the impartial culture model, empirical probabilities using a large database of real-world elections, and limiting probabilities in the three-candidate case for some minor cases left over from (Kamwa et al., 2023) under the impartial anonymous culture model.

\vspace{.1 in}

\textbf{Conflict of Interest and Funding Statement}: The authors have no relevant financial or non-financial interests to disclose. The authors did not receive support from any organization for the submitted work. \vspace{.1 in}

\textbf{Data Availability Statement}: Most of the data used in this article is publicly available at (Mattei \& Walsh, 2013), (Otis, 2022), and \url{https://github.com/mggg/scot-elex}. The rest is available upon request.

\section*{Appendix}\label{appendix}

We provide proofs of Proposition \ref{IC_prop} and Proposition \ref{IC_proppartial}. The proofs 
use geometric techniques developed in (Saari \& Tataru, 1999) and (Kamwa \& Merlin, 2015).
\vskip 2mm

\textbf{Proof of Proposition \ref{IC_prop}} (i)

\textbf{Step 1} First, we find the probability that  $A$ and $B$ are the plurality and IRV winners respectively. Following the arguments of  (Saari \& Tataru, 1999),  the probability that (\ref{oldorder}) and (\ref{oldwinner}) are satisfied is equal to  the area of the spherical simplex $S$  defined by these  inequalities on the surface of the unit sphere in $\mathbb{R}^3$, divided by the  area of this sphere. 
Let
 \begin{align*}
 \mathbf v_1 &= (1,1,-1,-1,0,0)\\
 \mathbf v_2 &= (0,0,1,1,-1,-1)\\
 \mathbf v_3 &= (-1,-1,1,1, -1, 1). 
 \end{align*}
  be the normal vectors of the three hyperplanes bounding $S$. By the Gauss-Bonet Theorem,  $Vol_2 (S) = \alpha_{12}+  \alpha_{13}+ \alpha_{23}-\pi$ where $\alpha_{ij}$ is the angle between vectors $\mathbf v_i$ and $\mathbf v_j$ and $Vol_n$ is the $n-$dimensional volume.
 
Using the law of cosines, 
 \begin{align*}
 \alpha_{12} &= \angle (\mathbf v_1, \mathbf v_2) = \cos^{-1}(\frac{-\mathbf v_1 \cdot \mathbf v_2}{\lVert \mathbf v_1 \rVert \lVert \mathbf v_2 \rVert} ) =   \cos^{-1}(\frac{2}{2 \cdot 2})= \cos^{-1}(\frac{1}{2}) \\
 \alpha_{13} &= \angle (\mathbf v_1, \mathbf v_3) = \cos^{-1}(\frac{-\mathbf v_1 \cdot \mathbf v_3}{\lVert \mathbf v_1 \rVert \lVert \mathbf v_3 \rVert} ) =   \cos^{-1}(\frac{4}{2 \cdot \sqrt{6}} )= \cos^{-1}(\frac{2}{\sqrt{6}}) \\ 
  \alpha_{23} &= \angle (\mathbf v_2, \mathbf v_3) = \cos^{-1}(\frac{-\mathbf v_2 \cdot \mathbf v_3}{\lVert \mathbf v_2 \rVert \lVert \mathbf v_3 \rVert} ) =   \cos^{-1}(\frac{-2}{2 \cdot \sqrt{6}}) = \cos^{-1}(\frac{-1}{\sqrt{6}} ).
 \end{align*} 
 
 Thus the probability is equal to 
 \begin{align*}
 \frac{1}{4 \pi}Vol_2(S) = \frac{1}{4 \pi}[ \alpha_{12}+\alpha_{12}+\alpha_{12} - \pi] = 0.04077671.
 \end{align*}
 
\vskip 3mm
 \textbf{Step 2}: Next, we find the probability that additionally (\ref{zcond}) is satisfied.
 Following (Kamwa \& Merlin, 2015), we introduce the  modified inequality with parameter $t \in [0,1],$
 \begin{equation}\label{modified}
 a_1+a_2>b_1+b_2+t(b_2-b_1).
 \end{equation}
Inequality (\ref{modified}) replaces (\ref{zcond}). If  $t=0$, (\ref{modified}) reduces to the first inequality in (\ref{oldorder});  if $t=1$, we recover (\ref{zcond}).  Similarly to Step 1, the probability that these conditions are met for a large number of voters $N$ is equal to the  volume of the spherical simplex $S$  defined by these  inequalities on the surface of the unit sphere in $\mathbb{R}^4$, divided by the volume of the surface of this sphere, $2\pi^2$. 
To find the $Vol_3(S)$, we use the Sch{\"a}fli formula 
 \begin{equation*}
d Vol_n = \frac{1}{n-1} \sum_{1 \le j < k \le n} Vol_{n-2}(S_j \cap S_k) d\alpha_{jk}
\end{equation*}
where $n=3$ and $S_i$ are the hyperplanes bounding $S$. Let $S_i$ be the hyperplane with normal vector $\mathbf v_i$.
Since (\ref{modified}) has normal vector $\mathbf v_4= (1,1,t-1,-t-1,0,0)$ and $\lVert \mathbf v_4 \rVert = \sqrt{4+2t^2}$, 
we have,  
  \begin{align*}
 \alpha_{14} &= \angle (\mathbf v_1, \mathbf v_4) =   \cos^{-1}(\frac{-4}{2 \cdot  \sqrt{4+2t^2}})= \cos^{-1}(\frac{-2}{\sqrt{4+2t^2}}) \\
 \alpha_{24} &= \angle (\mathbf v_2, \mathbf v_4) =   \cos^{-1}(\frac{2}{2 \cdot \sqrt{4+2t^2} } )= \cos^{-1}(\frac{1}{\sqrt{4+2t^2}}) \\ 
  \alpha_{34} &= \angle (\mathbf v_3, \mathbf v_4) =    \cos^{-1}(\frac{4}{\sqrt{6} \cdot \sqrt{4+2t^2}}).
   \end{align*} 
   Differentiating, gives
   \begin{align*}
d\alpha_{14} = \frac{-\sqrt{2}}{2+t^2} \quad  d\alpha_{24} = \frac{t}{\sqrt{3+2t^2} \cdot (2+t^2) }  \quad \mbox{and} \quad d\alpha_{34} =\frac{2t}{\sqrt{2+3t^2} \cdot  (2+t^2)}.
   \end{align*} 
  
  To calculate $Vol_1(S_j \cap S_k)$, we must determine where the hyperplanes intersect.  Let $P_{ijk}$ be the vertex lying at  the intersection of $S_i$, $S_j$ and $S_k$. We  identify a basis for the subspace orthogonal to that spanned by $\mathbf v_1 \ldots, \mathbf v_4$. One such basis is   $\mathbf v_5 = (1,1,0,0,0,0)$ and $\mathbf v_6 = (1,1,1,1,1,1).$

 Next, we find points $P_{ijk}$ lying at the intersection of the hyperplanes with normals given by the vectors $\mathbf v_i$, $\mathbf v_j$ and $\mathbf v_k$. Vertex $P_{124}$, for instance, can then be found by solving the conditions
   \begin{align*}
n_1+n_2-n_3-n_4& =0 \\
n_3+n_4-n_5-n_6 &=0 \\
-n_1-n_2+n_3+n_4-n_5+n_6 &>0 \\
n_1+n_2 +(t-1)n_3+(-t-1)n_4 &=0 \\
n_1-n_2 &=0 \\
n_1+n_2+n_3+n_4+n_5+n_6 &=0.
 \end{align*}
 This yields $P_{124}=(0,0,0, 0, -1, 1).$ Similarly, we have $P_{134}=(1,1,1,1,-2,-2)$ and $P_{234}=(1,1,\frac{-3-t}{2t}, \frac{3-t}{2t}, -2, 1).$
 
 Since $\lVert P_{124} \rVert  = \sqrt{2}$, $\lVert P_{134} \rVert  = 2\sqrt{3}$ and $\lVert P_{234} \rVert  =\frac{1}{t} \sqrt{\frac{3}{2}}\sqrt{5t^2+3}$, we have
 \begin{align*}
 Vol(S_1 \cap S_4)&= \angle (P_{124}, P_{134}) =\cos^{-1}(\frac{0}{\sqrt{2} \cdot 2 \sqrt{3}} ) =\frac{\pi}{2}\\
 Vol(S_2 \cap S_4)& = \angle (P_{124}, P_{234}) =\cos^{-1}(\frac{\sqrt{3}t}{\sqrt{5t^2+3}} ) \\
Vol(S_3\cap S_4)&= \angle (P_{134}, P_{234}) =\cos^{-1}(\frac{t}{\sqrt{2}\sqrt{5t^2+3}} ) 
 \end{align*}
By  the Sch{\"a}fli formula, 
$d Vol_3 = \frac{1}{2} \sum_{1 \le j < k \le 4} Vol_{1}(S_j \cap S_k) d\alpha_{jk}$, 
so
\begin{align*}
Vol_3(S_{t=1}) & = Vol_{3}(S_{t=0})+ \int_{0}^1 d Vol_3 S \\
& =   Vol_{3}(S_{t=0}) +\frac{1}{2} \int_{0}^1  Vol_{1}(S_1 \cap S_4) d\alpha_{14}+Vol_{1}(S_2 \cap S_4) d\alpha_{24}+Vol_{1}(S_3 \cap S_4) d\alpha_{34} dt \\
& =  Vol_{3}(S_{t=0}) +\frac{1}{2}\left[I_1+I_2+I_3 \right]
 \end{align*}
where 
$I_1 = \int^1_0 \frac{\pi}{2} \cdot \frac{-2}{4+2t^2} dt = -0.99679$
and similarly $I_2 =0.11231$ and $I_3 =0.312545$

By Step 1,  $Vol_{3}(S_{t=0}) = Vol_2(S_{t=0})\frac{2\pi^2}{4\pi} =0.04077671 (2\pi^2)$. Hence the probability that all these conditions are met is equal to
\begin{equation*}
\frac{1}{2\pi^2} \left[0.04077671 (2\pi^2) + \frac{1}{2}(-0.99679+  0.11231+ 0.312545) \right] =0.026289.
\end{equation*}

Since the choice of  $A$ and $B$ as plurality and IRV winners was arbitrary, the  probability of a negative participation paradox is $6 \times 0.026289 \approx 0.157737.$

\textbf{Proof of (ii)} Since a negative participation paradox  can only occur when the IRV and plurality winners are distinct, the probability that an election
 demonstrates a negative participation paradox given that these winners are different is equal to the probability of negative participation paradox divided by the probability that the IRV and plurality winners are distinct.  The latter  was determined in Step 1; hence this probability is equal to $0.026289/0.04077671 \approx 0.644717.$ $\qed$

 \vskip 2mm

The proof of Proposition \ref{IC_proppartial} is similar to that of Proposition \ref{IC_prop}.

\vskip 2mm
\textbf{Proof of Proposition \ref{IC_proppartial} (i)}

\textbf{Step 1} First, we find the probability that  $A$ and $B$ are the plurality and IRV winners respectively.
The inequalities (\ref{partial_oldorder}) and (\ref{partial_oldwinner}) have normal vectors
 \begin{align*}
 \mathbf v_1 &= (1,1,1,-1,-1,-1,0,0,0)\\
 \mathbf v_2 &= (0,0,0,1,1,1,-1,-1,-1)\\
 \mathbf v_3 &= (-1,-1,-1,1,1,1,0,-1,1). 
 \end{align*}
So \begin{align*}
 \alpha_{12} &= \angle (\mathbf v_1, \mathbf v_2) = \cos^{-1}(\frac{-\mathbf v_1 \cdot \mathbf v_2}{\lVert \mathbf v_1 \rVert \lVert \mathbf v_2 \rVert} ) =   \cos^{-1}(\frac{3}{\sqrt{6} \cdot \sqrt{6} })= \cos^{-1}(\frac{1}{2}) \\
 \alpha_{13} &= \angle (\mathbf v_1, \mathbf v_3) = \cos^{-1}(\frac{-\mathbf v_1 \cdot \mathbf v_3}{\lVert \mathbf v_1 \rVert \lVert \mathbf v_3 \rVert} ) =   \cos^{-1}(\frac{6}{\sqrt{6} \cdot \sqrt{8}} )= \cos^{-1}(\frac{\sqrt{3}}{2}) \\ 
  \alpha_{23} &= \angle (\mathbf v_2, \mathbf v_3) = \cos^{-1}(\frac{-\mathbf v_2 \cdot \mathbf v_3}{\lVert \mathbf v_2 \rVert \lVert \mathbf v_3 \rVert} ) =   \cos^{-1}(\frac{-3}{\sqrt{6} \cdot \sqrt{8}}) = \cos^{-1}(\frac{-\sqrt{3}}{4} ).
 \end{align*} 
 
 Thus, the probability is equal to 
 \begin{align*}
 \frac{1}{4 \pi}Vol_2(S) = \frac{1}{4 \pi}[ \alpha_{12}+\alpha_{12}+\alpha_{12} - \pi] =0.03563737.
 \end{align*}

\vskip 3mm
 \textbf{Step 2}: Next, we find the probability that additionally (\ref{partial_zcond}) is satisfied. We introduce the inequality $a_0+a_1+a_2 >b_0+b_1+b_2+ t(b_2-b_1)$. When $t=0$ this reduces to a (\ref{partial_oldorder});  when $t=1$ we get the required extra condition.

Since this inequality has normal vector equal to
$ \mathbf v_4 = (1,1,1,-1,t-1,-t-1,0,0,0)$, we have

 \begin{align*}
 \alpha_{14} &= \angle (\mathbf v_1, \mathbf v_4) =   \cos^{-1}(\frac{-6}{\sqrt{6} \cdot \sqrt{6+2t^2}})= \cos^{-1}(\frac{-\sqrt{3}}{\sqrt{3+t^2}})\\
 \alpha_{24} &= \angle (\mathbf v_2, \mathbf v_4) =  \cos^{-1}(\frac{3}{\sqrt{6} \cdot \sqrt{6+2t^2}}) = \cos^{-1}(\frac{\sqrt{3}}{2\sqrt{3+t^2}})\\ 
  \alpha_{34} &= \angle (\mathbf v_3, \mathbf v_4) =   \cos^{-1}(\frac{6}{\sqrt{8} \cdot \sqrt{6+2t^2}}) = \cos^{-1}(\frac{3}{2\sqrt{3+t^2}})
   \end{align*} 
leading to 

  \begin{align*}
d\alpha_{14} = \frac{-\sqrt{3}}{(3+t^2)} \quad  d\alpha_{24} = \frac{\sqrt{3}t}{\sqrt{9+4t^2}(3+t^2)}  \quad \mbox{and} \quad d\alpha_{34} =\frac{3t}{\sqrt{3+4t^2}(3+t^2)}.
   \end{align*} 

Next, we require a basis for the space in $\mathbb{R}^9$ that is orthogonal to $\mathbf v_1, \mathbf v_2, \mathbf v_3$ and $\mathbf v_4.$ 
One such basis is 
 \begin{align*}
 \mathbf v_5&= (1,1,-2,0,0,0,0,0,0)\\
 \mathbf v_6 &= (1,-1, 0,0,0,0,0,0,0)\\
  \mathbf v_7&= (0,0,0,-2,1,1,0,0,0)\\
 \mathbf v_8&= (0,0,0,0,0,0,-2,1,1)\\
\mathbf v_9 & = (1,1,1,1,1,1,1,1,1).
  \end{align*}

We find $P_{124}$, $P_{134}$ and $P_{234}$ as before to get $P_{124}=(0,0,0, 0,0,0, 0, -1, 1),$  $P_{134}=(1,1,1,1,1,1,-2,-2,-2)$ and $P_{234}=(4t,4t,4t,-2t,-9-2t,9-2t,-2t,-11t,7t).$
 Since $\lVert P_{124} \rVert  = \sqrt{2}$, $\lVert P_{134} \rVert  = 3\sqrt{2}$ and $\lVert P_{234} \rVert  = \sqrt{48t^2+8t^2+121t^2+49t^2+162+8t^2} =\sqrt{234t^2+162} =3\sqrt{2(13t^2+9)}$, we have

 \begin{align*}
 Vol(S_1 \cap S_4)&= \angle (P_{124}, P_{134}) =\cos^{-1}(\frac{0}{\sqrt{2} \cdot 2 \sqrt{18}} ) =\frac{\pi}{2}\\
 Vol(S_2 \cap S_4)& = \angle (P_{124}, P_{234}) =\cos^{-1}(\frac{18t}{\sqrt{2} \cdot 3\sqrt{2(13t^2+9)}} )= \cos^{-1}(\frac{3t}{\sqrt{13t^2+9}})\\
Vol(S_3\cap S_4)&= \angle (P_{134}, P_{234}) =\cos^{-1}(\frac{18t}{3\sqrt{2} \cdot 3\sqrt{2(13t^2+9)}} )= \cos^{-1}(\frac{t}{\sqrt{13t^2+9}})
 \end{align*}

So \begin{align*}
Vol_3(S_{t=1}) & = Vol_{3}(S_{t=0})+ \int_{0}^1 d Vol_3 S \\
& =   Vol_{3}(S_{t=0}) +\frac{1}{2} \int_{0}^1  Vol_{1}(S_1 \cap S_4) d\alpha_{14}+Vol_{1}(S_2 \cap S_4) d\alpha_{24}+Vol_{1}(S_3 \cap S_4) d\alpha_{34} dt \\
& =  Vol_{3}(S_{t=0}) +\frac{1}{2}\left[I_1+I_2+I_3 \right]
 \end{align*}
where 
$I_1 = \int^1_0 \frac{\pi}{2} \cdot  \frac{-\sqrt{3}}{(3+t^2)} dt = -0.82247$
and similarly $I_2 =0.0806918$ and $I_3 =0.28144178$

By Step 1,  $Vol_{3}(S_{t=0}) = Vol_2(S_{t=0})\frac{2\pi^2}{4\pi} =0.03563737(2\pi^2)$. Hence the probability that all these conditions are met is equal to
\begin{equation*}
\frac{1}{2\pi^2} \left[0.03563737 (2\pi^2) + \frac{1}{2}( -0.82247+ 0.0806918+ 0.28144178) \right] =0.023976912.
\end{equation*}

Since the choice of $A$ and $B$ was arbitrary, the probability  of a negative participation paradox is  is $6 \times 0.023976912 \approx 0.143861$

\textbf{Proof of (ii)}
As with the case with complete ballots,  the probability that an election demonstrates a negative participation paradox given that these winners are different is equal to the probability of negative participation paradox divided by the probability that the IRV and plurality winners are distinct. The latter was determined in Step 1; hence this probability is equal to $0.023976912/0.03563737 \approx 0.672803. \qed$

\end{document}